\documentclass[acmsmall]{acmart}
\usepackage{multirow}
\usepackage[normalem]{ulem}
\graphicspath{ {./Figs/} }
\usepackage{hyperref}
\hypersetup{hidelinks,
	colorlinks=true,
	allcolors=black,
	pdfstartview=Fit,
	breaklinks=true}
\usepackage{amsmath}
\usepackage{booktabs} % For formal tables
\usepackage{graphicx}
\usepackage{balance}
\usepackage{epstopdf}
\usepackage{subfigure}
\usepackage{diagbox}
\usepackage{multirow}
\usepackage{makecell}
\usepackage{color}
\usepackage{colortbl}
\usepackage[ruled,vlined,linesnumbered]{algorithm2e}
\usepackage{bbm}
\usepackage{extarrows}

\usepackage{array}

\AtBeginDocument{%
	}

\setcopyright{acmlicensed}
\copyrightyear{2025}
\acmYear{2025}
\begin{document}
\title{Discrete Scale-invariant Metric Learning for Efficient Collaborative Filtering}
%\author{Yan Zhang$^{\dagger}$, Hongzhi Yin$^{\ddagger}$, Zi Huang$^{\ddagger}$, Xingzhong Du$^{\ddagger}$, Guowu Yang$^{\dagger}$, Defu Lian$^{\dagger}$}
%\affiliation{%
%  \institution{$^{\dagger}$School of Computer Science and Engineering, University of Electronic Science and Technology of China \\
%  $^{\ddagger}$School of Information Technology and Electrical Engineering, The University of Queensland}
%  \email{yixianqianzy@gmail.com}

%\footnote{$^{\star}$This author is the corresponding author.}
%\author{Yan Zhang$^{\dagger, \ddagger}$, Hongzhi Yin$^{\ddagger},^*$}
%\author{Yan Zhang$^{\dagger, \ddagger}$, Hongzhi Yin$^{\ddagger}$\footnotemark[1], Zi Huang$^{\ddagger}$, Xingzhong Du$^{\ddagger}$, Guowu Yang$^{\dagger}$, Defu Lian$^{\dagger}$}
%\renewcommand{\thefootnote}{\fnsymbol{footnote}}
%
%\affiliation{%
%  \institution{$^{\dagger}$School of Computer Science and Engineering, University of Electronic Science and Technology of China\\
%  $^{\ddagger}$School of Information Technology and Electrical Engineering, The University of Queensland}
%}
%\email{yixianqianzy@gmail.com, h.yin1@uq.edu.au, huang@itee.uq.edu.au, x.du@uq.edu.au, guowu@uestc.edu.cn, dove.ustc@gmail.com}
%\renewcommand{\shortauthors}{Y. Zhang et al.}
% \author{Anonymity}

%
%
\author{Yan Zhang}
\authornote{Corresponding author.}
\email{yan.zhang@cdu.edu.au}
\affiliation{%
	\institution{Faculty of Science and Technology, Charles Darwin University}
	% \streetaddress{2006 Xiyuan Avenue}
	%  \city{Chengdu}
	%  \state{Sichuan}
	\country{Darwin, Australia}
	% \postcode{611731}
}

\author{Li Deng}
\email{136995963@qq.com }
\affiliation{%
	\institution{University of Electronic Science and Technology of China}
	% \streetaddress{0810 Ellengowan Dr, Casuarina}
	%  \city{Darwin}
	%  \state{Northern Territory (NT)}
	\country{Chengdu, China}}

\author{Lixin Duan}
\email{lxduan@uestc.edu.cn}
\affiliation{%
	\institution{Shenzhen Institute for Advanced Study, University of Electronic Science and Technology of China}
	% \streetaddress{2006 Xiyuan Avenue}
	%  \city{Shenzhen}
	%  \state{Guangdong}
	\country{Shenzhen, China}
	% \postcode{518100}
}

\author{Sami Azam}
\email{sami.azam@cdu.edu.au}
\affiliation{%
	\institution{Faculty of Science and Technology, Charles Darwin University}
% \streetaddress{2006 Xiyuan Avenue}
%  \city{Chengdu}
%  \state{Sichuan}
\country{Darwin, Australia}}

% \IEEEcompsocitemizethanks{
% \IEEEcompsocthanksitem Y. Zhang and Sami Azam are with the IT Discipline, Faculty of Science and Technology, Charles Darwin University, \protect\\ E-mail: yan.zhang@cdu.edu.au. 

% <-this % stops an unwanted space
%\thanks{Manuscript received June 30, 2018.}
% }

% \IEEEtitleabstractindextext{%
% \begin{abstract}

%\author{Hongzhi Yin}
%\authornote{This is the corresponding author}
%\affiliation{%
%  \institution{School of Information Technology and Electrical Engineering, The University of Queensland}
%}
%\email{h.yin1@uq.edu.au}
%
%\author{Zi Huang, Xingzhong Du}
%\affiliation{%
%  \institution{School of Information Technology and Electrical Engineering, The University of Queensland}
%}
%\email{huang@itee.uq.edu.au, x.du@uq.edu.au}
%
%\author{Guowu Yang, Defu Lian}
%\affiliation{%
%  \institution{School of Computer Science and Engineering, University of Electronic Science and Technology of China}
%}
%\email{guowu@uestc.edu.cn, dove.ustc@gmail.com}

%\titlenote{Produces the permission block, and
% copyright information}
%\subtitle{Extended Abstract}
%\subtitlenote{The full version of the author's guide is available as
%  \texttt{acmart.pdf} document}

\begin{abstract}
%\footnotetext[1]{The corresponding authors}
Metric learning has attracted extensive interest for its ability to provide personalized recommendations based on the importance of observed user-item interactions. Current metric learning methods aim to push negative items away from the corresponding users and positive items by an absolute geometrical distance margin. However, items may come from imbalanced categories with different intra-class variations. Thus, the absolute distance margin may not be ideal for estimating the difference between user preferences over imbalanced items. To this end, we propose a new method, named discrete scale-invariant metric learning (DSIML), by adding binary constraints to users and items, which maps users and items into binary codes of a shared Hamming subspace to speed up the online recommendation. Specifically, we firstly propose a scale-invariant margin based on angles at the negative item points in the shared Hamming subspace. Then, we derive a scale-invariant triple hinge loss based on the margin. To capture more preference difference information, we integrate a pairwise ranking loss into the scale-invariant loss in the proposed model. Due to the difficulty of directly optimizing the mixed integer optimization problem formulated with \textit{log-sum-exp} functions, we seek to optimize its variational quadratic upper bound and learn hash codes with an alternating optimization strategy. Experiments on benchmark datasets clearly show that our proposed method is superior to competitive metric learning and hashing-based baselines for recommender systems.

\end{abstract}


% \begin{IEEEkeywords}
% Recommender system, metric learning, hash codes, ranking
% \end{IEEEkeywords}}

\maketitle

\section{Introduction} \label{sec1}

Personalized recommender systems, as one of the most critical and effective approaches for alleviating information overload, have been successfully applied in various areas including online e-commerce websites: Amazon, Netflix, Yelp, etc., online educational systems, and even online health-care systems. Personalized recommender systems often provide top-$k$ ranked list of items for a specific user to assist users to target their interests. As implicit feedback information is more common than explicit rating data in the real world applications, we focus on recommendations based on the implicit feedback data in this paper. 

Efforts have been made to improve the accuracy of personalized recommendations by estimating the preference difference between positive and negative items based on implicit feedback (e.g., clicks, browse recodes, rating records, etc.). These efforts are two folds: (1) ranking-based recommender systems consisting of pairwise learning to rank (LTR) \cite{rendle2009bpr,rendle2014improving,li2021new,moffat2012models,wu2023graph,hu2022artificial} and listwise LTR methods \cite{weimer2008cofi,srebro2004maximum,wu2018sql}; and (2) metric learning recommender systems \cite{hsieh2017collaborative,li2020symmetric,ma2020probabilistic}.  

%Despite that, ranking-based recommendations are powerful for learning the user preference difference by learning discriminative latent factors for users and items by pairwise ranking losses and listwise ranking objectives. However, they suffer from a crucial drawback stemming from MF-based methods that violate the triangle inequality according to \cite{shrivastava2014asymmetric}, which leads to the failure of capturing more preference information \cite{hsieh2017collaborative}. To address the limitations of MF-based methods, metric learning approaches have been studied in recommendations \cite{hsieh2017collaborative,li2020symmetric,ma2020probabilistic}. These methods project users and items into a low-dimension metric space, where the distance estimates the user preference for the item. Nevertheless, the existing metric learning method has the limitation of assuming all users and items share the same fixed margin $m$, which is inappropriate for personalized recommendations. 

%There are two ways to estimate the user preference difference between positive items and negative items based on implicit feedbacks: (1) ranking-based recommender systems consisting of pairwise learning to rank (LTR) and listwise LTR methods; and (2) metric learning recommender systems. 
In pairwise LTR methods, one representative is Bayesian personalize ranking (BPR) \cite{rendle2009bpr} and its variant \cite{rendle2014improving} based on a non-uniform item sampler, which is proposed to speed up the convergence under the circumstance that item popularity has a tailed distribution. To achieve better ranking performances, nBRP-MF \cite{li2021new} formulates ranking tasks of recommendations based on the Rank-Biased Precision(BRP) \cite{moffat2012models}. It evaluates that both pairwise and listwise ranking-based recommendation models can achieve better performances for active users with more of the relevant interactions compared to less active users. To address the ranking problem for binary relevance datasets, collaborative less-is-more filtering (CLiMF) \cite{shi2012climf} is proposed by directly maximizing the Mean Reciprocal Rank (MRR). To achieve efficient recommendations with a simpler student model, a ranking distillation method \cite{tang2018ranking,zhang2019efficient} is proposed for learning to rank problems by training a teacher-student model. 

In listwise LTR methods, one representative is CoFi-Rank \cite{weimer2008cofi}, which optimizes a listwise ranking-based objective motivated by the idea of maximum margin matrix factorization (MMMF) \cite{srebro2004maximum}. Then, Yue et. al \cite{shi2010list} proposes the ListRank-MF for collaborative filtering that combines a listwise LTR objective with matrix factorization. A ranked list of items is obtained by minimizing a loss function that represents the uncertainty between training lists and output lists produced by a MF ranking model. Following the memory-based collaborative filtering (CF) framework, Shanshan et al \cite{huang2015listwise} proposes the ListCF that predicts a total order of items for each user based on similar users' probability distributions over permutations of the items. One important advantage of ListCF lies in its ability of reducing the computational complexity of the training and prediction procedures while achieving the same or better ranking performances as compared to previous ranking-oriented memory-based CF algorithms. To relieve the expensive computational cost and provide a theoretical analysis for the listwise ranking model, SQL-Rank \cite{wu2018sql} is proposed by taking listwise collaborative ranking as maximum likelihood under a permutation model which applies probability mass to permutations based on a low-rank latent rating matrix. 

Despite that, ranking-based recommendations are powerful for learning the user preference difference by learning discriminative latent factors for users and items by pairwise ranking losses and listwise ranking objectives. However, they suffer from a crucial drawback stemming from MF-based methods that violate the triangle inequality according to \cite{shrivastava2014asymmetric}, which leads to the failure of capturing more preference information \cite{hsieh2017collaborative}. To address the limitations of MF-based methods, metric learning approaches have been studied in recommendations \cite{hsieh2017collaborative,li2020symmetric,ma2020probabilistic,xu2022metacar,zhang2022diverse}. These methods project users and items into a low-dimension metric space, where the distance estimates the user preference for the item. Nevertheless, the existing metric learning method has the limitation of assuming all users and items share the same fixed margin $m$, which is inappropriate for personalized recommendations. 

\begin{figure}[!tb]
	\centering
	\includegraphics[width=0.6\linewidth]{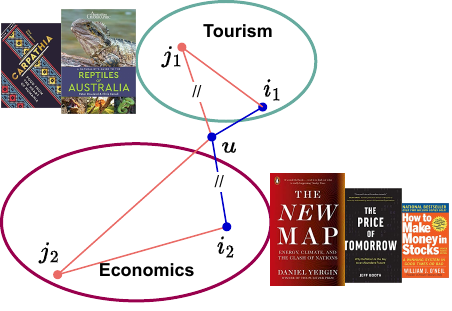}
	%		\subfigure{\includegraphics[width=1\linewidth]{mzf}}
	\caption{An illustration of the challenge brought by imbalanced items. User $u$ has positive items $i_{1}$, $i_{2}$ and negative items $j_{1}$, $j_{2}$ from two imbalanced classes of books: the majority Economics and the minority Tourism. An absolute distance margin is difficult to measure the user preference difference between $i_{2}$ and $j_{1}$ when $d(u, j_{1})= d(u, i_2)$.}\label{fig:cha2}
\end{figure}

Accordingly, it is challenging for existing metric learning methods to learn the user preference difference when facing imbalance data shown in Figure \ref{fig:cha2}. It shows two classes of books on Amazon: Economics and Tourism. Economics is the majority class, which is much more diverse than the minority class Tourism books, and the scale of Economics books is much larger than Tourism books. Therefore, it's not applicable to fix the same margin for the imbalanced item dataset. For user $u$, the positive (interested) items are $i_{1}, i_{2}$, the negative (not interested) items are $j_{1}, j_{2}$. Existing metric learning methods push the negative item $j_{1}$ away from user $u$ with an absolute geometrical distance margin $m$ when training on the triplets $(u, i_{1}, j_{1})$ and $(u, i_{2}, j_{2})$; meanwhile, it will push the positive item $i_{2}$ away from user $u$ because $d(u, j_{1})= d(u, i_2)$, which hinders to learn the user preference difference between items $j_{1}$ and $i_{2}$. So a geometrical distance margin is not feasible to estimate the user preference difference to items from imbalanced classes.

\section{The Proposed Model} \label{dsml-model}
In this section, we first introduce the notations used in this Chapter. Next, we intuitively analyze the relation of a user, her/his positive item, and the negative item of a triplet in Hamming space. Then, we introduce the model formulation of DSIML in Section \ref{dsml-model-formul}. This section first proposes a scale-invariant margin for metric learning in Hamming space. To capture more preference information, we integrate a pairwise ranking loss with the scale-invariant loss in the proposed model DSIML. 

%%\vspace{-6pt}
\subsection{Notations} \label{dsml-model-nota}
Let users and items be denoted as $\mathcal{U}=\{1,\cdots ,n\} $ and $\mathcal{I}=\{1\cdots ,m\}$ respectively. The implicit feedback $ \mathcal{S}$ is defined as a subset of the cartesian product of $\mathcal{U}$ and $\mathcal{I}$: $\mathcal{S}\subseteq \mathcal{U}\times \mathcal{I}$, where $s_{ui}= 1$ denotes user $u$ has an interaction/rating with item $i$; otherwise, $s_{ui} = 0$. The task of personalized top-$k$ recommendation is providing a personalized ranking list of $k$ items from all items in $\mathcal{I}$ for a specific user. The proposed method aims to provide personalized recommendations with metric learning in Hamming space by projecting users and items into $d$-dimension binary codes. 
I define positive items of user $i$ as $\mathcal{I}_{u}^{+}=\left\{ i\in \mathcal{I}:\left( u,i \right)\in \mathcal{S} \right\}\subset \mathcal{I}$; and negative items as the remaining items $\mathcal{I}_{u}^{-}=\mathcal{I}\backslash \mathcal{I}_{u}^{+} \subset \mathcal{I}$. Similarly, we define positive users who rated item $i$ as $\mathcal{U}_{i}^{+}=\left\{ u\in \mathcal{U}:\left( u,i \right)\in \mathcal{S} \right\}\subset \mathcal{U}$; and negative users as other users $\mathcal{U}_{i}^{-}=\mathcal{U}\backslash \mathcal{U}_{i}^{+} \subset \mathcal{U}$.
\subsection{Model Formulation} \label{dsml-model-formul}
Suppose that we are given a set of training triplets $\Omega=\{(u, i, j): u\in \mathcal{U}, i\in \mathcal{I}_{u}^{+}, j\in \mathcal{I}_{u}^{-}\}$. In existing metric learning methods, such as CML \cite{hsieh2017collaborative}, it aims to push negative item $j$ away from user $u$ by a distance margin $m>0$ compared to the positive item $i$ by minimizing the following relaxed surrogate hinge loss,
\begin{equation} \label{cml}
\begin{gathered}
\mathcal{L}_{triplet}=\big[d^{2}(\mathbf{b}_{u}, \mathbf{d}_{i}) - d^2(\mathbf{b}_{u}, \mathbf{d}_{j})+m\big]_{+}.
\end{gathered}
\end{equation}
where $[\cdot]_{+}=\max\{0, \cdot\}$ represents the hinge function. $\mathbf{b}_{u}$ denotes the representation of user $i$. $\mathbf{d}_{i}, \mathbf{d}_{j}$ represent vectors of positive item $i$ and negative item $j$, respectively. 

As analyzed in SML \cite{li2020symmetric} and PALAM \cite{ma2020probabilistic}, CML only considers user-item relations and ignores the item-item relations, so it probably drags the negative item towards the positive item, which contradicts the objective of metric learning. Hence, SML and PALAM are proposed by considering user-item and item-item relations.  

However, SML and PALAM cannot deal with the case in Figure~\ref{fig:cha2} because the absolute distance margin cannot be used to measure the user preference difference for items from imbalanced categories. In this paper, we design a scale-invariant margin to deal with the imbalanced problem in Hamming space. Suppose that $\mathbf{b}_{u}\in \mathbf{B}$ denotes the hash code of user $u$. $\mathbf{d}_{i}\in \mathbf{D}$ denotes the hash code of item $j$. The distance between user $i$ and item $j$ in Hamming space is denoted as $\text{Dis}_{H}(\mathbf{b}_{u}, \mathbf{d}_{i})$. In the triplet $(u,i,j)\in \Omega$, if $\text{Dis}_{H}(\mathbf{b}_{u}, \mathbf{d}_{i}) < \text{Dis}_{H}(\mathbf{b}_{u}, \mathbf{d}_{j})$, then user $u$ is more interested in item $i$ than item $j$. The Hamming distance $\text{Dis}_{H}$ can be denoted by the inner product of two hash codes, 
%\begin{equation}\label{eq:pre}
\begin{align}
\text{Dis}_{H}(\mathbf{b}_{u}, \mathbf{d}_{i}) 
&=\sum\limits_{k=1}^{d}{\mathbb{I}\left( {{b}_{uk}}\neq{{d}_{ik}} \right)} \nonumber\\
&=\frac{1}{2}(d - \mathbf{b}_{u}^{T}{{\mathbf{d}}_{i}}) \nonumber\\
&=\frac{1}{2}(d - ||\mathbf{b}_{u}||\cdot||{\mathbf{d}}_{i}|\cos\theta_{ui}),
\end{align}
%\end{equation}
where $d$ is the dimension of hash codes $\mathbf{b}_{u}$ and $\mathbf{d}_{i}$. $\theta_{ui}$ denotes the angle formed by the two binary vectors. From the above equation, we derive that the Hamming distance is consistent with the cosine value of the angle, because the norm of $d$-dimension hash code is constant $\sqrt{d}$. Consequently, a smaller angle leads to similar hash codes, which motivates us to find a scale-invariant margin based on the angles of hash codes.

\begin{figure}[!htbp]
	\centering
	\includegraphics[width=0.6\linewidth]{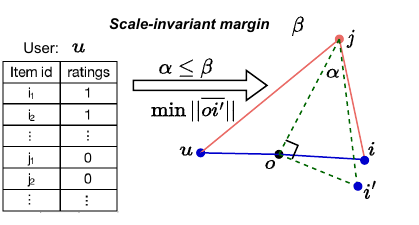}
	%		\subfigure{\includegraphics[width=1\linewidth]{mzf}}
	\caption{The illustration of deriving scale-invariant margin. If I directly reduce the angle $\angle uij$, the negative point $j$ may be dragged towards the positive point $i$, which violates the objective. Alternatively, we reduce $\alpha$ in the auxiliary triangle $\triangle oi'j$ with the intuition of pushing negative items away from an approximated local distribution formed by $u$ and $i$. Hence, we impose a scale-invariant margin $\beta$ on $\alpha$, i.e., $\alpha \le \beta$.}\label{fig:dsml-margin}
	%%\vspace{-1em}
\end{figure}

As illustrated in Figure \ref{fig:dsml-margin}, the objective \eqref{cml} of CML aims to minimize the edge $\overline{ui}$ and push the negative sample $j$ away from $u$ with a geometrical margin $m$, that enforces $||\overline{ui}||+m\le ||\overline{uj}||$, which may drag the negative item $j$ towards to the positive item $i$. To avoid this, we penalize a longer edge $\overline{ij}$ compared to the edge $\overline{ui}$. According to the cosine rule, it can be proved that the angle surrounded by the longer edge $\overline{ui}$ and $\overline{ij}$ should be the smallest one in the triangle $\triangle uij$ \cite{wang2017deep}. As the summarization of three angles in the triangle equals $180^{\circ}$, so the angle $\angle uji$ should be smaller than $60^{\circ}$, i.e., $\angle uji \le 60^{\circ}$. Since items may belong to different categories with different intra-class variations, we aim to find an angle margin $\beta$ imposed on the angle at the negative item point $j$ instead of searching for an absolute distance margin $m$. 

Given user $u$, the positive item $i$, and the negative item $j$, the proposed objective is to minimize $\angle uji$. If we decrease the angle $\angle uji$, the negative $j$ may be pulled towards the positive $i$, which violates the proposed objective. So we try to find an alternative way to push the negative item $j$ away from both the user and the positive item. The intuition is pushing negative sample $j$ away from a local distribution formed by user $u$ and the positive item $i$. Approximately, we assume the local distribution is the circle area built by the diameter $\overline{ui}$. Thus when we move the user $u$ to the center $o$ and move the positive item $i$ to the item $i'$. $||\overline{oi'} || = ||\overline{oi} || = \frac{1}{2}||\overline{ui} ||$, and $\overline{oi'}$ is orthogonal to $\overline{oj}$. Given user $u$ and positive item $i$, if we reduce the angle $\alpha$, the negative item $j$ will be pushed away from all points in the local distribution, including $u$, $i$, $o$, and $i'$. This motivates us to define an angle margin $\beta$ on the angle $\alpha$, and the proposed objective is 
\begin{equation}
\alpha \le \beta.
\end{equation}
In the triangle $\triangle oi'j$, 
\begin{equation}
\tan {\alpha} = \frac{||\overline{oi'}||}{||\overline{oj}||} \le \tan{\beta}.
\end{equation}
In the training procedure on the triplet $\{u,i,j\}$, we aim to update hash codes $\mathbf{b}_{u}, \mathbf{d}_{i}$, and $\mathbf{d}_{j}$ by minimizing the following hinge loss:
\begin{equation} \label{hinge}
\mathcal{L}_{scale} = \big[||\overline{oi'}||^2 - \gamma^{2}||\overline{oj}||^2\big]_{+},
\end{equation}
where $\gamma = \tan{\beta}$, $||\overline{oi'}||^{2}=||\overline{oi}||^{2}=\frac{1}{4}(\mathbf{b}_{u}-\mathbf{d}_{i})^{2}$. $||\overline{oj}||^2=||\mathbf{d}_{j}-\frac{1}{2}(\mathbf{b}_{u}+\mathbf{d}_{i})||^2$. 

If $\alpha \le \beta$, i.e., $||\overline{oi'}||^2 \le \gamma^{2}||\overline{oj}||$, we do not update hash codes $\mathbf{b}_{u}, \mathbf{d}_{i}$, and $\mathbf{d}_{j}$; otherwise, we update them by minimizing the following scale-invariant margin loss:
\begin{equation}
\mathcal{L}_{scale} = ||\mathbf{b}_{u}-\mathbf{d}_{i}||^2-\gamma^{2}||2\mathbf{d}_{j}-(\mathbf{b}_{u}+\mathbf{d}_{i})||^2.
\end{equation}
To better understand the effect of optimizing the scale-invariant margin loss, we investigate sub-problems regarding $\mathbf{b}_{u}$, $\mathbf{d}_{i}$ and $\mathbf{d}_{j}$, respectively. By dropping terms irrelevant to $\mathbf{b}_{u}$ when $\mathbf{d}_{i}$ and $\mathbf{d}_{j}$ are fixed, the sub-problem of $\mathbf{b}_{u}$ is as follows,
\begin{equation}
\mathcal{L}_{scale-u} = -(1+\gamma^2)\mathbf{b}_{u}^{T}\mathbf{d}_{i}+2\gamma^2\mathbf{b}_{u}^{T}\mathbf{d}_{j},
\end{equation}
we derive that the updates rule of $\mathbf{b}_{u}$ is dependent on all samples in the triplet $(u,i,j)\in \Omega$. Similarly, we derive the sub-problem of $\mathbf{d}_{i}$ and $\mathbf{d}_{j}$ as follows,
\begin{align}
\mathcal{L}_{scale-i} = &-(1+\gamma^2)\mathbf{b}_{u}^{T}\mathbf{d}_{i}+2\gamma^2\mathbf{d}_{i}^{T}\mathbf{d}_{j},\\
\mathcal{L}_{scale-j} =& 2\gamma^2(\mathbf{b}_{u}+\mathbf{d}_{i})^{T}\mathbf{d}_{j}.
\end{align}
Compared to CML, where each sample in a triplet is updated with only two points \cite{li2020symmetric}, which may easily stick into a locally optimal solution. In the proposed scale-invariant metric learning, each sample $u$, $i$, and $j$ is updated by all three points. So it can achieve a better solution than CML.

Compared to SML and PALAM, the proposed scale-invariant margin is enforced on the angle at the negative point. The absolute distance margin of SML or PALAM varies with items from classes with different intra-class variations. Instead, the proposed margin does not change with different variations. So the margin is invariant for items with imbalanced classes.

However, the hinge loss \eqref{hinge} is a non-smooth objective when considering more than one negative point. So we convert the hinge loss to its smooth upper bound inspired by the recent work \cite{sohn2016improved}. Because users and items are of constant length, i.e., $||\mathbf{b}_{u}||=||\mathbf{d}_{i} || = ||\mathbf{d}_{j}||=\sqrt{d}$ in Equation\eqref{hinge}. After dropping constant terms, Equation~\eqref{hinge} is approximated by the following smooth \textit{log-sum-up} upper bound,
\begin{equation}\label{scale}
\mathcal{L}_{scale}=\sum_{(u,i,j)\in\Omega}\log\big(1+\exp(2\gamma^{2}(\mathbf{b}_{u}^{T}\mathbf{d}_{i}+\mathbf{d}_{i}^{T}\mathbf{d}_{j})- 
(1+\gamma^2)\mathbf{b}_{u}^{T}\mathbf{d}_{i})\big), 
\end{equation}
To capture more preference information from triplets $(u,i,j)\in \Omega$, we add the pairwise ranking loss to the objective inspired by BPR \cite{rendle2009bpr}. In Hamming space, we assume that the rating score is predicted by $\hat{r}_{ui} = 1-\frac{1}{d}\text{Dis}_{H}(\mathbf{b}_{u}, \mathbf{d}_{i})= \frac{1}{2}+\frac{1}{2d}\mathbf{b}_{u}^{T} \mathbf{d}_{i}$. So the pairwise ranking loss is formulated as:
\begin{equation}
\mathcal{L}_{pair} = \sum_{(u,i,j)\Omega}\mathbb{I}(\hat{r}_{ui}-\hat{r}_{uj}),
\end{equation}
where $\mathbb{I}(\cdot)$ is a delta function, which returns 1 if the input is true and 0 otherwise. Optimizing the above objective often directly leads to an NP-hard problem~\cite{gao2013one}. A solvable solution is to minimize some pairwise surrogate losses as follows,
\begin{equation}
\mathcal{L}_{pair}= \sum_{(u,i,j)\in\Omega} \ell({{{\hat{r}}}_{ui}} - {{{\hat{r}}}_{uj}})
\end{equation}
where $\ell: \mathbb{R} \rightarrow \mathbb{R}^+$ is the surrogate \textit{log-sum-exp} function \cite{zhang2017discrete} as follows
\begin{equation}\label{pair}
\mathcal{L}_{pair}= \sum_{(u,i,j)\in\Omega} \log(1+\exp(\frac{\mathbf{b}_{u}^{T}\mathbf{d}_{j} -\mathbf{b}_{u}^{T}\mathbf{d}_{i}}{2d} )).
\end{equation}
Optimizing the above objective can capture more user preference difference information between positive and negative items.

By combining the pairwise ranking loss \eqref{pair} and the scale-invariant loss \eqref{scale}, we formulate DSIML as the following objective,
\begin{equation}\label{siml}
\min_{\mathbf{B},\mathbf{D}}\mathcal{L}_{\tiny{dsiml}} (\mathbf{B},\mathbf{D})= \mathcal{L}_{pair} + \lambda \mathcal{L}_{scale},
\end{equation}
where $\lambda$ is a trade-off hyper-parameter that weights the importance between the pairwise loss and the scale-invariant loss. In the experiments, we test the effect of $\lambda$ on recommendation performances. Its continuous version of DSIML, referred to as SIML has the same objective of dropping the binary constraints.
%\subsection{The Objective of DSIML}\label{sec33}
%To obtain hash codes of users and items, we
\section{Model Optimization} \label{dsml-opt}
Due to the intractability of solving the objective \eqref{siml} with binary constraints imposed on $\mathbf{B}$ and $\mathbf{D}$, it's hard to find a closed form updates rule. Inspired by the new discrete optimization method \cite{zhang2018discrete}, we seek to minimize a variational quadratic upper bound of the objective of DSIML in Equation~\eqref{siml} based on \textit{JJ bound} \cite{jaakkola1997variational}. The upper bound of \textit{log-sum-exp} function is as follows
\begin{equation}\label{lse}
\log\big(1+e^{t}\big) \le \pi(\xi)(t^2 - \xi^2)+ \frac{1}{2} (t - \xi) + \log\big(1+e^{\xi}),
\end{equation}
where $\pi(\xi) = \frac{1}{2\xi}\big(\sigma(\xi) - \frac{1}{2}\big)$, $\sigma(x) = \frac{1}{1+e^{-x}}$ is a sigmoid function. we call $\xi$ as the variational parameter. It's evident that the upper bound is tight at $\xi=t$. By denoting $\hat{\ell}(t, \xi)=\pi(\xi)(t^2 - \xi^2)+ \frac{1}{2} (t - \xi) + \log\big(1+e^{\xi}\big)$, we derive the following upper bound of Equation~\eqref{siml},
\begin{equation}\label{scale_up}
\begin{aligned}
&\min_{\mathbf{B},\mathbf{D},\Phi,\mathcal{H}} \hat{\mathcal{L}}_{\tiny{dsiml}}=\sum_{(u,i,j)\in\Omega}\big(\hat{\ell}(x_{uij}, \varphi_{uij}) +\lambda \hat{\ell}(y_{uij}, \eta_{uij})\big),\\
&\text{s.t.} \ \ \mathbf{B}\in\{\pm 1\}^{n\times d},  \mathbf{D}\in\{\pm 1\}^{m\times d}, \Phi \in \mathbb{R}^{|\Omega|}, \mathcal{H} \in \mathbb{R}^{|\Omega|}.
\end{aligned}
\end{equation}
where $x_{uij}=\frac{1}{2d}(\mathbf{b}_{u}^{T}\mathbf{d}_{j} -\mathbf{b}_{u}^{T}\mathbf{d}_{i})$ is from Equation~\eqref{pair}, and $y_{uij}=2\gamma^{2}(\mathbf{b}_{u}^{T}\mathbf{d}_{j}+\mathbf{d}_{i}^{T}\mathbf{d}_{j})-(1+\gamma^2)\mathbf{b}_{u}^{T}\mathbf{d}_{i}$ is from Equation~\eqref{scale}. $\varphi_{uij}\in \Phi$ and $\eta_{uij}\in \mathcal{H}$ are the variational parameters.

To optimize the above mixed integer problem, we adopt an alternating optimization method, which is commonly applied in solving mixed integer problems \cite{zhang2016discrete, zhang2017discrete}. 
\paragraph{Updating users' hash codes $\mathbf{B}$.}
Given $\mathbf{D}$, $\Phi$, and $\mathcal{H}$, we update $\mathbf{b}_{u}$ of each user $u$ in parallel. By dropping irrelevant items to $\mathbf{b}_{u}$, the sub-problem is a binary quadratic problem (BQP) with respect to $\mathbf{b}_{u}$,
\begin{equation}\label{u-obj}
\min_{\mathbf{b}_{u}\{\pm 1\}^{d}} \mathbf{b}_{u}^{T}\mathbf{A}_{u} \mathbf{b}_{u}+ \mathbf{c}_{u}^{T}\mathbf{b}_{u},
\end{equation}
where  $\mathbf{A}_{u} = \mathbf{A}_{xu} + 4\lambda d^2 \mathbf{A}_{yu}$,  $\mathbf{c}_{u} = \mathbf{c}_{xu} + 4\lambda d^2 \mathbf{c}_{yu}$. $\mathbf{A}_{xu}$ and $\mathbf{c}_{xu}$ come from the loss $\hat{\ell}(x_{uij}, \varphi_{uij})$, and $\mathbf{A}_{yu}$ and $\mathbf{c}_{yu}$ come from the loss $\hat{\ell}(y_{uij}, \eta_{uij})$, and
\begin{align}\label{cdu}
\begin{cases}
\mathbf{A}_{xu}=&\sum_{(i,j)}\big( \mathbf{d}_{i}\mathbf{d}_{i}^{T}\pi(\varphi_{uij})+\mathbf{d}_{j}\mathbf{d}_{j}^{T}\pi(\varphi_{uij})-2\mathbf{d}_{j}\mathbf{d}_{i}^{T}\pi(\varphi_{uij})\big);  \\
\mathbf{A}_{yu}=&4\gamma^4\sum_{(i,j)}\big( 4(1+\gamma^2)^2\mathbf{d}_{i}\mathbf{d}_{i}^{T}\pi(\eta_{uij}) + \mathbf{d}_{j}\mathbf{d}_{j}^{T}\pi(\eta_{uij})- \\
&(1+\gamma^2)\mathbf{d}_{j}\mathbf{d}_{i}^{T}\pi(\eta_{uij})\big).  \\
\mathbf{c}_{xu}=&\sum_{(i,j)} d(\mathbf{d}_{j}-\mathbf{d}_{i}); \\
\mathbf{c}_{yu}=&\sum_{(i,j)} \big(4\gamma^4(\mathbf{d}_{i}^{T}\mathbf{d}_{j})\mathbf{d}_{j}\pi(\eta_{uij})- 8\gamma^4(1+\gamma^2)(\mathbf{d}_{i}^T\mathbf{d}_{j})\mathbf{d}_{i} + \\
&2\gamma^2\mathbf{d}_{j} - (1+\gamma^2)\mathbf{d}_{i}\big).
%\vspace{-10pt}
\end{cases}
\end{align}
Previous studies mainly apply the Discrete coordinate descent(DCD) \cite{Shen_2015_CVPR} method to learn hash codes in a bitwise way. As analyzed in DRMF  \cite{zhang2018discrete}, the discrete coordinate descent method easily falls into local optima and is very sensitive to the initialization of parameters. In contrast to discrete coordinate descent, we update the entire hash code ${{\mathbf{b}}_{u}}$ in each iteration by directly optimizing the BQP problem with the BQP solver \footnote{https://www.cvxpy.org}. In this paper, we choose the Xpress optimizer to optimize the above BQP problem by transferring hash codes $\mathbf{b}_{u} \in \{\pm 1\}^{d}$ to boolean vectors $\frac{1}{2}(\mathbf{b}_{u}+\mathbf{1}_{d})\in \{0,1\}^{d}$.

\paragraph{Updating items' hash codes $\mathbf{D}$.}
Similarly, given $\mathbf{B}$, $\Phi$, and $\mathcal{H}$, we update $\mathbf{d}_{j}$ of each item $j$ in parallel. By dropping irrelevant items to $\mathbf{d}_{j}$, the sub-problem is a binary quadratic problem (BQP) with respect to $\mathbf{d}_{j}$,
\begin{equation}\label{v-obj}
\min_{\mathbf{d}_{i}\{\pm 1\}^{d}} \mathbf{d}_{i}^{T}\mathbf{A}_{i} \mathbf{d}_{i}+ \mathbf{c}_{i}^{T}\mathbf{d}_{i},
\end{equation}
where  $\mathbf{A}_{i} = \mathbf{A}_{xi} + 4\lambda d^2 \mathbf{A}_{yi}$,  $\mathbf{c}_{i} = \mathbf{c}_{xi} + 4\lambda d^2 \mathbf{c}_{yi}$. Similarly, $\mathbf{A}_{xu}$ and $\mathbf{c}_{xu}$ come from the loss $\hat{\ell}(x_{uij}, \varphi_{uij})$, and $\mathbf{A}_{yu}$ and $\mathbf{c}_{yu}$ come from the loss $\hat{\ell}(y_{uij}, \eta_{uij})$, and 
\begin{align}\label{cdj}
\begin{cases}
\mathbf{A}_{xu}=&\sum_(u,j) \mathbf{b}_{u}\mathbf{b}_{u}^{T}\pi(\varphi_{uij}); \\
\mathbf{A}_{yj} = &\sum_(u,j)  \big(4\gamma^4\mathbf{d}_{j}\mathbf{d}_{j}^{T}\pi(\eta_{uij})+ (1+\gamma^2)^2 \mathbf{b}_{u}\mathbf{b}_{u}^{T}\pi(\eta_{uij}) - \\
&4\gamma^2(1+\gamma^2) \mathbf{d}_{j}\mathbf{b}_{u}^{T}\pi(\eta_{uij})  \big).\\
\mathbf{c}_{xi}=&\sum_{(u,i)} \big(-2(\mathbf{b}_{u}^{T}\mathbf{d}_{j})\mathbf{b}_{u}\pi(\varphi_{uij}) - d\mathbf{b}_{u} \big); \\
\mathbf{c}_{yu}=&\sum_{(u,j)} \big(4\gamma^4(\mathbf{b}_{u}^{T}\mathbf{d}_{j})\mathbf{d}_{j}\pi(\eta_{uij}) - \\ &2\gamma^2(1+\gamma^2)(\mathbf{b}_{u}^{T}\mathbf{d}_{j})\mathbf{b}_{u}\pi(\eta_{uij}) + 
2\gamma^2 \mathbf{d}_{j} - (1+\gamma^2) \mathbf{b}_{u} \big).
\end{cases}
\end{align}
Similarly, we optimize the above BQP with the assist of Xpress solver by transferring hash codes $\mathbf{d}_{i} \in \{\pm 1\}^{d}$ to boolean vectors $\frac{1}{2}(\mathbf{d}_{i}+\mathbf{1}_{d})\in \{0,1\}^{d}$.

\paragraph{Updating variational parameters $\Phi$.}
Given $\mathbf{B}$, $\mathbf{D}$, and $\mathcal{H}$, we obtain the following sub-objective of $\varphi_{uij}$ by discarding irrespective terms of $\varphi_{uij}$,
\begin{equation}\label{phi_obj}
\min_{\varphi_{uij}\in \mathbb{R}} \pi(\varphi_{uij})(x_{uij}^2 - \varphi_{uij}^2)+ \frac{1}{2} (x_{uij} - \varphi_{uij}) + \log\big(1+e^{\varphi_{uij}})
\end{equation}
The sub-problem \eqref{phi_obj} is convex since its second derivative is greater than zero. Hence, the optimal value is achieved when the derivation of Equation~\eqref{phi_obj} equals zero. For each user, the derivation of Equation~\eqref{phi_obj} equals zero at
\begin{equation}\label{phi}
{{\varphi}_{uij}}=x_{uij}=\frac{1}{2d}(\mathbf{b}_{u}^{T}\mathbf{d}_{j} -\mathbf{b}_{u}^{T}\mathbf{d}_{i})  
\end{equation}
In practical, we don't need store $\{{\varphi}_{uij}: \forall (u,i,j)\in \Omega\}$ in the training procedure. we only need to store all expressions $x_{uij}$ and then update ${{\varphi}_{uij}}$. we do also not need to store $\pi({{\varphi}_{uij}})$, since it can be computed on-the-fly in constant time when we cache these $x_{uij}$.

\paragraph{Updating variational parameters $\mathcal{H}$.}
Similarly, given $\mathbf{B}$, $\mathbf{D}$, and $\Phi$, we derive the following sub-problem of $\eta_{uij}$ by dropping irrelevant terms of $\eta_{uij}$ in Equation\eqref{phi_obj},
\begin{equation}\label{eta_obj}
\min_{\eta_{uij}\in \mathbb{R}} \pi(\eta_{uij})(y_{uij}^2 - \eta_{uij}^2)+ \frac{1}{2} (y_{uij} - \eta_{uij}) + \log\big(1+e^{\eta_{uij}})
\end{equation}
Similarly, the sub-objective \eqref{eta_obj} is convex and the optimal value is achieved when the derivation of Equation~ \eqref{eta_obj} equals to zero. For each user, the derivation of Equation~ \eqref{eta_obj} equals to zero at
\begin{equation}\label{eta}
{{\eta}_{uij}}=y_{uij}=2\gamma^{2}(\mathbf{b}_{u}^{T}\mathbf{d}_{i}+\mathbf{d}_{i}^{T}\mathbf{d}_{j})-(1+\gamma^2)\mathbf{b}_{u}^{T}\mathbf{d}_{i} 
\end{equation}
Similarly, we don't need store ${\eta}_{uij}$ and $\pi({{\eta}_{uij}})$, and we only need to store all expressions $y_{uij}$ and then update ${{\eta}_{uij}}$ and $\pi({{\eta}_{uij}})$ with these $y_{uij}$. Finally, we summarize the optimization steps in Algorithm \ref{alg:dsml}. 

\begin{algorithm}
	\KwIn{User-item implicit feedback $\mathcal{S}$, dimension $d=20$, $\gamma=1$, $\lambda=1$.}
	\KwOut{Hash codes of users and items $\mathbf{B}$ and $\mathbf{D}$.}
	%\caption{Discrete Ranking-based Matrix Factorization (DRMF)\label{alg:drmf}} % $\mathbf{P},{Q}\leftarrow$DRMF($\mathbf{R}, \mathbf{w}, \mathbf{P}_0,\mathbf{Q}_0$)}
	\caption{$\text{DSIML}$}\label{alg:dsml} % $\mathbf{P},\mathbf{Q}\leftarrow$DRMF($\mathbf{R}, \mathbf{w}, \mathbf{P}_0,\mathbf{Q}_0$)}
	\textbf{Initialize:} Firstly we learn continuous vectors $\mathbf{U}$ and $\mathbf{V}$ by solving Equation~ \eqref{siml} without binary constraints, then we initialize $\mathbf{B}\leftarrow sign(\mathbf{U})$, $\mathbf{D}\leftarrow sign(\mathbf{V})$,
	$\Phi \leftarrow Equation~\eqref{phi}$, $\mathcal{H} \leftarrow Equation~\eqref{eta}$\;
	%$\mathbf{w} = \mathbf{1}^{|S|}$\;
	\Repeat{$\mathcal{L}_{dsml}$ is convergent}	
	{	
		\For{$i \in \{1,\cdots, n\}$}
		{
			\For{$j \in \mathcal{I}_{u}^{+}$ and $k \in I_{u}^{-}$}{
				Compute $\varphi_{uij}$ by Equation~\eqref{phi} \; 
			}
			Update $\mathbf{b}_u$ by sloving the objective \eqref{u-obj} by Xpress solver\;
		}
		
		\For{$i\in \{1,\cdots,m\}$}
		{
			\For{$u \in \mathcal{U}_{i}^{+}$ and $k \in I_{u}^{-}$}{
				Compute $\eta_{uij}$ by Equation~\eqref{eta}  \; 
			}
			Update $\mathbf{d}_i$ by sloving the objective \eqref{v-obj} by Xpress solver\;
		}
	}
\end{algorithm}

In the training stage, we optimize the objective \eqref{siml} by a mini-batch containing multiple triplets from the same users. Specifically, we divide triplets, w.r.t users. Suppose the batch size is 16, we train the proposed model on all triplets of 16 users. In training, we optimize user by user, i.e., we train the proposed model in the way of N-pair sampling \cite{sohn2016improved}. It considers an ($N+1$)-tuplet of training examples $\{u, i, j_{1}, j_{2}, \dots, j_{N-1}\}$ by sampling $N-1$ negative samples for each positive feedback $s_{ij}=1$. As evaluated in NCF \cite{he2017neural}, the recommender system can achieve good performance when we randomly sample four to eight negative samples for each positive feedback. So in this Chapter, we sample five negative samples for each positive item to train the proposed model.
%%\vspace{-6pt}
%\subsection{Relation to Other Models} \label{sec_related}
%drawbacks
%1. item-item relation
%2. fixed m, is difficult to optimize for all users and items
%3. Margin based on Euclidean distance is not appropriate for different categories of items.
%CML, 
%SML, 
%PALAM
%\begin{table}
%\caption{Relation to Other Models}
%\small
%\centering
%%\begin{tabular*}{0.41\paperwidth}{@{\extracolsep{\fill}}ccccccc}
%\begin{tabular}{c|c|c|c}
%	\hline
%	%  \hline
%	Methods & Q1&Q2 & Q3\\
%	\hline
%	CML& - & -  & -\\
%	\hline
%	PALAM& \checkmark  & \checkmark  & -  \\
%	\hline
%	SML &  \checkmark &\checkmark  &  - \\
%		SIML&  \checkmark &\checkmark  &  \checkmark 	  \\
%	\hline
%\end{tabular}
%\label{tab2}
%\end{table}

\section{Experiments}\label{dsml-exp}
This section puts forward a new metric learning method SIML (continuous version of DSIML). In this section, we evaluate the performance of the hashing-based recommendation framework DSIML by comparing it with the competitive hashing-based recommendations DCF, DPR, CCCF, LightRec, and the latest PHD \cite{hansen2021projected}. Besides, we evaluate the effectiveness of the proposed metric learning method SIML(continuous version of DSIML) by comparing with state-of-the-are metric learning recommendation methods CML \cite{hsieh2017collaborative}, SML \cite{li2020symmetric}, one representative pairwise ranking-based recommendation framework BPR \cite{rendle2009bpr}, the competitive baseline LightGCN \cite{he2020lightgcn}, and the latest ranking-based recommendation method CPR \cite{wan2022cross}. 

%we do not compare the proposed method with the latest competitive collaborative filtering method LightGCN \cite{he2020lightgcn} because it exploits both user-item interactions and user-user relations by a second-order embedding smoothness step. While, the proposed method and baselines only apply user-item interactions for recommendations. So it's not fair to compare the proposed method with LightGCN. 

we aim to answer the following questions to evaluate the effectiveness of the proposed method:
%is proposing the scale-invariant metric learning hashing-based recommendation framework, which can capture more preference difference between negative items and positive items than existing metric learning methods. 
\begin{itemize}
	\item \textbf{RQ1:} Does the proposed DSIML improve the recommendation performance compared with hashing-based recommender systems?
	\item \textbf{RQ2:} Does the proposed metric learning with scale-invariant margin improve the recommendation performance compared with state-of-the-art metric learning methods for recommendations?
	\item \textbf{RQ3:} Does the proposed DSIML speed up the online recommendation compared with one continuous representative method BPR?
	\item \textbf{RQ4:} Does the scale-invariant margin effectively cope with the imbalanced data problem?	
	\item \textbf{RQ5:} How do the hyper-parameters $\lambda$ influence the recommendation performance?
\end{itemize}
%%\vspace{-4pt}
\subsection{Experimental Settings} \label{dsml-exp-set}
\subsubsection{Datasets}
we evaluate the performance of the proposed method and baselines with three real-world datasets: CDs, Movies, and Books, which are subset datasets of Amazon\footnote{http://jmcauley.ucsd.edu/data/amazon/}. Amazon datasets cover user interactions on items and item content on 24 product categories. For datasets we used in this section, we remove users and items with less than 20 ratings. Then, we convert explicit ratings into implicit feedback, where `1' represents a user rated an item that is considered as a positive item. Items with a value `0' are considered as negative items. The statistics of datasets are listed in Table \ref{tab:dsml-data}.

\begin{table}[!htbp]
	%\extrarowheight=3pt
	%\vspace{-8pt}
	\caption{Statistics of datasets.}
	%	\small
	%\vspace{-8pt}
	\centering
	%\begin{tabular*}{0.41\paperwidth}{@{\extracolsep{\fill}}ccccccc}
	\begin{tabular}{c|c|c|c|c}
		\toprule
		%  \hline
		Dataset & Users & Items& Ratings & Sparsity\\
		CDs &   25,400&24,904 & 43,903 & 99.99\%  \\
		\hline
		Movies &  18,128&11,252 & 784,382 & 99.62\%  \\
		\hline
		Books   & 603,374   &348,957  & 8,575,000  & 99.99\% \\	
		%		\hline
		%    	Yelp& 50,059 & 61,634 & 1,005,090 & 99.96\%  \\
		\toprule	
	\end{tabular}
	%\vspace{-10pt}
	\label{tab:dsml-data}
\end{table}
we randomly split positive implicit feedback into 80\% for training and the remaining 20\% for testing. For each positive item, we randomly sample 5 negative items for training inspired by NCF \cite{he2017neural}. 80\% positive ratings together with these sampled negative ratings make up the training dataset. we report the performance of top-$k$ recommendations under two metrics: \textit{normalized discounted cumulative gain} ($NDCG@k$) \cite{jarvelin2017ir} and \textit{hit ratio} ($HR@k$) \cite{yin2013lcars} since they have been widely used for evaluating the performance of ranking tasks.
%\paragraph{Evaluation Protocols}
%Following the common strategy of implicit feedbacks \cite{he2017neural}, I adopt the similar \textit{leave-one-out} evaluation protocols for evaluating the performance of the proposed method and other baselines. Specifically, I sample 99 negative items for each positive item to evaluate the proposed DSIML and other baselines. 20\% positive ratings together with these 99 sampled negative ratings make up the testing data $D_{test}$. Then I report the performance of top-$k$ recommendations under two metrics: \textit{normalized discounted cumulative gain} (NDCG) \cite{jarvelin2017ir} and \textit{hit ratio} (HR) \cite{yin2013lcars} since they have been widely used for evaluating ranking tasks, especially for recommender systems.

\subsubsection{Hyper-parameter Settings}
we tune the optimal hyper-parameters for the proposed method by the grid search. Specifically, we search the hyper-parameter $\lambda$ for each dataset within $\{1e^{-3}, 1e^{-2}, 1e^{-1}, 1e^{1}, 1e^{2}\}$. For the scale-invariant margin $\gamma=\tan \beta$, due to the limitation of the smallest angle $\alpha \le 60^{\circ}$ in the triangle $\triangle oi'j$, and $\alpha \le \beta$, so we tune the margin $\gamma$ from $\{0.2, 0.5, \dots, 1.7\}$ due to $\tan(60^{\circ})\approx 1.7$. we report the results of tuning the two hyper-parameters in the following experiments on Movies
%%\vspace{-6pt}
%\begin{figure*}[!tb]
%	\centering
%	\includegraphics[width=0.95\linewidth]{fig11}
%	%\vspace{-10pt}
%	\caption{The recommendation performance comparison of hashing-based methods under $NDCG@k$ on CDs, Movies, and Books.}\label{fig:cmp-hash}
%	%\vspace{-6pt}
%\end{figure*}

\subsubsection{Baselines} 
As introduced above, to evaluate the accuracy improvement of the proposed discrete method DSIML, we compare the performance of DSIML with the state-of-the-art hashing-based methods DCF, CCCF, LightRec, and PHD. Besides, to evaluate the effectiveness of metric learning proposed in this chapter, we compare the performance of the continuous version of SIML with the state-of-the-art five continuous frameworks BPR, CML, SML, LightGCN, and CPR.
\begin{itemize}
	\item \textbf{DCF:} Discrete collaborative filtering \cite{zhang2016discrete} is the state-of-the-art hashing-based recommender system, which directly learns hash codes by optimizing a discrete optimization problem. 
	%	\item \textbf{DPR:} Discrete pairwise ranking \cite{zhang2017discrete} is formulated as a pairwise ranking objective. DPR also learns hash codes by optimizing a mixed integer optimization problem directly.
	\item \textbf{CCCF:} Compositional coding for collaborative filtering \cite{liu2019compositional} constructs a compositional hashing method with $G$ components for each user and item and optimizes a discrete problem directly.
	\item \textbf{LightRec:} LightRec \cite{lian2020lightrec} is a lightweight recommender system which is constituted by a backbone of $B$ codebooks, each of which is composed of $W$ latent codewords.
	\item \textbf{PHD:} Hansen et.al puts forward a Variational Hashing (VH) model \cite{hansen2021projected} that mainly composed of a new measurement of dissimilarity, Projected Hamming Dissimilarity (PHD) between items in Hamming subspace. These dissimilarities are considered as the important weights of bits in hash codes. 
	\item \textbf{BPR:} Bayesian Personalized Ranking framework~\cite{rendle2009bpr} is a ranking-based recommendation based on Matrix Factorization, which directly optimizes ranking based evaluation metrics with Bayesian. 
	%	\item \textbf{MBRP-MF:} nBRP-MF\citep{li2021new} is an optimized ranking based recommendation technique which is inspired by BRP. MBRP-MF evaluates that ranking-based recommendation models can achieve better ranking performance for active users with more of relevant interactions compared to less active users. It is evaluated a promising alternative when learning to rank in the recommendation context.
	\item \textbf{CML} Collaborative metric learning \cite{hsieh2017collaborative} learns a joint metric space to encode user-user relations together with relations of user and positive item. It is the first work that combines metric learning into recommendations.	
	\item \textbf{SML} Symmetric metric learning \cite{li2020symmetric} aims to search different optimal margin for each user and each item with a symmetric hinge loss.
	\item \textbf{LightGCN} LightGCN \cite{he2020lightgcn} constructs Graph Convolution Network (GCN) based on user-item interactive data, and exploits both user-item interactions and user-user relations by a second-order embedding smoothness step.
	\item \textbf{CPR} Cross Pairwise Ranking (CPR) proposes an unbiased recommendation by changing point-wise and pairwise rating loss.
\end{itemize}
\begin{figure*}[!htbp]
	\centering
	\includegraphics[width=1\linewidth]{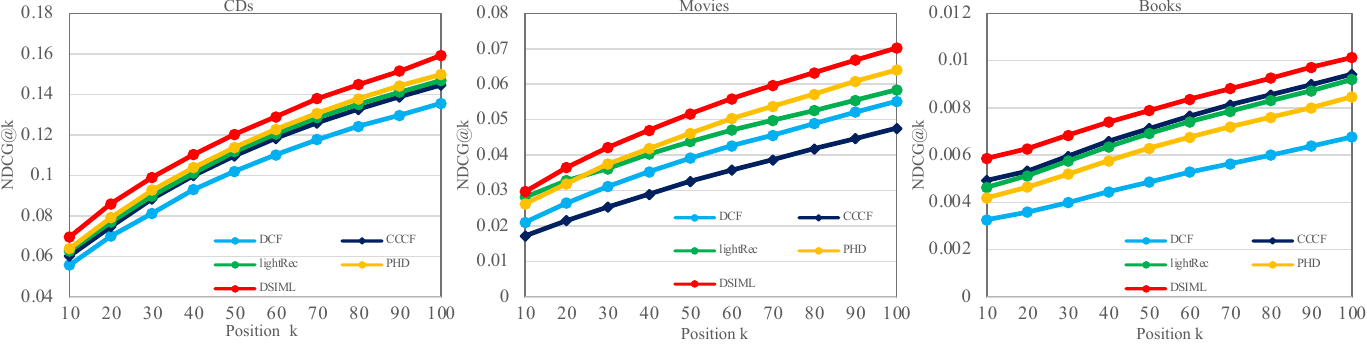}
	%\vspace{-6pt}
	\caption{The performance of the proposed discrete DSIML compared with state-of-the-art discrete recommendation methods on CDs, Movies, and Books under NDCG@k.}\label{fig:cmp-hash}
	\vspace{-10pt}
\end{figure*}
\begin{figure}[!htbp]
	\centering
	\includegraphics[width=1\linewidth]{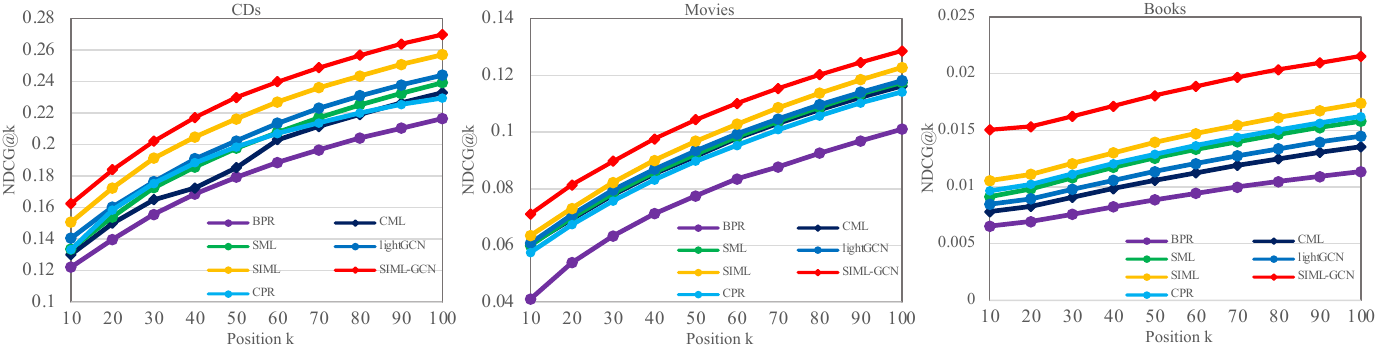}
	%\vspace{-6pt}
	\caption{The performance of the proposed continuous SIML and SIML-GCN (the variant on SIML) compared with sate-of-the-art metric learning based recommendation models and other competitive continuous recommendation methods on CDs, Movies and Books under NDCG@k.}\label{fig:cmp-real}
	\vspace{-10pt}
\end{figure}
\begin{table*}[!tb] 
	%\extrarowheight=3pt
	\caption{The performance of DSIML compared with state-of-the-art discrete methods on CDs, Movies and Books under $HR@k$. The best performance is highlighted with bold text.} \label{tab:cmp-hash}
	%\begin{tabular*}{0.41\paperwidth}{@{\extracolsep{\fill}}ccccccc}
	%		\linespread{1.5}
	%		\small
	\vspace{-8pt}
	\centering
	\begin{tabular}{c||ccc||ccc }
		\toprule
		%		   \hlinefill
		Dataset&\multicolumn{3}{c||}{CDs}&\multicolumn{3}{c}{Movies}\\
		\hline 
		Metric & $HR@10$&$HR@50$ & $HR@100$ &$HR@10$& $HR@50$& $HR@100$\\
		\hline \hline
		DCF & 0.0504 &	0.1768 	&0.2821 	&0.0175 	&0.0680 	&0.1184	\\
		\cline{2-7}
		CCCF & 	0.0543 	&0.0191 	&0.0301 	&0.0142 &	0.0567 &	0.1043  \\
		\cline{2-7}
		LightRec &	0.0562 &	0.1936 &	0.3038 &	0.0226 &	0.0705 	&0.1160  \\
		\cline{2-7}
		PHD&0.0578 	&0.1965 	&0.3095 	&0.0213 	&0.0774 &	0.1332  \\	
		\cline{2-7}
		\textbf{DSIML}&	\textbf{0.0621} 	&\textbf{0.2038 }&	\textbf{0.3207} 	&\textbf{0.0236}& 	\textbf{0.0859 }	&\textbf{0.1439 }	 \\
		%			\hline 
		%	\textit{Imp.}(\%) & 0.1032 	0.3939 	0.6026 	0.0738 	0.2672 	0.4055 	0.0833 	0.3144 	0.3839  \\		
		\toprule
	\end{tabular}
	\centering
	\begin{tabular}{c||cccccc }
		\toprule
		Dataset & \multicolumn{6}{c}{Books}\\
		\hline 
		Metric &$HR@10$ & $HR@30$ & $HR@50$&$HR@70$&$HR@90$&$HR@100$ \\
		\hline
		DCF &0.0019 &	0.0045 	&0.0068& 	0.0088 &	0.0109 	&0.0119 	\\
		\cline{2-7}
		CCCF & 0.0028 	&0.0066 &	0.0098 &	0.0124 &	0.0147 &	0.0159  \\
		\cline{2-7}
		LightRec &0.0025 &	0.0063 	&0.0093 	&0.0118 &	0.0137 	&0.0145  \\
		\cline{2-7}
		PHD &0.0024 &	0.0059 &	0.0087 &	0.0112 &	0.0134 	&0.0148  \\
		\cline{2-7}
		\textbf{DSIML} 	&\textbf{0.0034 }&	\textbf{0.0074} &	\textbf{0.0102} &	\textbf{0.0127}& \textbf{0.0152 }	&\textbf{0.0163}   \\
		%	\hline 
		%	\textit{Imp.}(\%) & 0.1032 	0.3939 	0.6026 	0.0738 	0.2672 	0.4055 	0.0833 	0.3144 	0.3839  \\		
		\toprule
	\end{tabular}
\end{table*}
\subsection{Experimental Results and Analysis}\label{dsml-exp-result}
In this section, we evaluate the performance of DSIML by comparing it with competitive hashing-based baselines and metric learning methods by answering five questions proposed at the beginning of Section \ref{dsml-exp}.
\subsubsection{The Performance Compared with Hashing-based Methods(RQ1) }
As shown in Figure \ref{fig:cmp-hash} and Table \ref{tab:cmp-hash}, the performances under two metrics, NDCG@k and HR@k, of the proposed DSIML are superior to other hashing-based baselines, which indicates that the proposed scale-invariant metric learning is effective to capture users' preferences over items from unbalanced classes. Besides, it also evaluates that the combination of a pairwise ranking loss and scale-invariant metric learning loss effectively provides accurate recommendations. we formulate DSIML with the ranking loss instead of a rating loss used in DCF. That's why DSIML performs better than DCF. In addition, LightRec, CCCF, and PHD are formulated with rating loss objectives other than the ranking loss objective used in DSIML. That's why they perform inferior to DSIML.

\begin{table*}[!tb] 
	%\extrarowheight=3pt
	\caption{The performance of continous SIML and SIML-GCN (the variant on SIML) compared with sate-of-the-art metric learning based recommendation models and other competitive continuous recommendation methods on CDs, Movies and Books under under $HR@k$. The best performance is highlighted with bold text.} \label{tab:cmp-real}
	%\begin{tabular*}{0.41\paperwidth}{@{\extracolsep{\fill}}ccccccc}
	%		\linespread{1.5}
	%		\small
	\vspace{-8pt}
	\centering
	\begin{tabular}{c||ccc||ccc }
		\toprule
		%		   \hlinefill
		Dataset&\multicolumn{3}{c||}{CDs}&\multicolumn{3}{c}{Movies}\\
		\hline 
		Metric & $HR@10$&$HR@50$ & $HR@100$ &$HR@10$& $HR@50$& $HR@100$\\
		\hline \hline
		BPR & 0.1013 	&0.2768 	&0.4042 	&0.0368 	&0.1187 	&0.1854\\
		\cline{2-7}
		CML & 	0.1113 	&0.2828 	&0.4243 	&0.0470 	&0.1427 	&0.2223   \\
		\cline{2-7}
		SML &	0.1145& 	0.2970 	&0.4283 &	0.0461 &	0.1405 	&0.2183  \\
		\cline{2-7}
		LightGCN&0.1186 	&0.3047 &	0.4357 &	0.0476 	&0.1452 	&0.2260  \\	
		\cline{2-7}
		CPR&0.1157& 	0.2992 	&0.4314 &	0.0468 	&0.1440 &	0.2216   \\	
		\cline{2-7}
		\textbf{SIML}&\textbf{0.1281} 	&\textbf{0.3220 }	&\textbf{0.4501 }	&\textbf{0.0505 }	&\textbf{0.1520} &\textbf{	0.2346} 	\\
		\cline{2-7} 
		\textbf{SIML-GCN} & \textbf{0.1352} &\textbf{	0.3380 }&	\textbf{0.4643 }	&\textbf{0.0547} &	\textbf{0.1628} 	&\textbf{0.2477  }   \\
		%			\hline 
		%	\textit{Imp.}(\%) & 0.1032 	0.3939 	0.6026 	0.0738 	0.2672 	0.4055 	0.0833 	0.3144 	0.3839  \\		
		\toprule
	\end{tabular}
	\centering
	\begin{tabular}{c||cccccc }
		\toprule
		Dataset & \multicolumn{6}{c}{Books}\\
		\hline 
		Metric &$HR@10$ & $HR@30$ & $HR@50$&$HR@70$&$HR@90$&$HR@100$ \\
		\hline
		BPR&0.0036 	&0.0082 &	0.0117 	&0.0148 	&0.0174 	&0.0186	\\
		\cline{2-7}
		CML &0.0044 	&0.0098 	&0.0140 &	0.0176 	&0.0208 	&0.0221  \\
		\cline{2-7}
		SML &0.0052 	&0.0115 	&0.0162 &	0.0202 	&0.0237 &	0.0252   \\
		\cline{2-7}
		LightGCN &0.0048 &	0.0105 &	0.0149 &	0.0187 	&0.0220 	&0.0235   \\
		\cline{2-7}
		CPR	&0.0054 	&0.0118 	&0.0167 	&0.0209 &	0.0243 	&0.0260 \\
		\cline{2-7}
		\textbf{SIML} &\textbf{0.0049 }	&\textbf{0.0118 }	&\textbf{0.0171 }	&\textbf{0.0212 }	&\textbf{0.0249 }	&\textbf{0.0266 }	 \\
		\cline{2-7}
		\textbf{ SIML-GCN}&\textbf{	0.0141 }	&\textbf{0.0191} &	\textbf{0.0232 }	&\textbf{0.0268} 	&\textbf{0.0298} 	&\textbf{0.0312 }\\
		%	\hline &
		%	\textit{Imp.}(\%) & 0.1032 	0.3939 	0.6026 	0.0738 	0.2672 	0.4055 	0.0833 	0.3144 	0.3839  \\		
		\toprule
	\end{tabular}
\end{table*}
\subsubsection{The Performance Compared with Metric Learning Methods(RQ2) }
As shown in Figure \ref{fig:cmp-real} and Table \ref{tab:cmp-real}, they reflect the performance of the proposed SIML and its invariant SIML-GCN under two metrics $NDCG@k$ and $HR@k$ are significantly superior to two other competing metric learning based recommendation models, CML and SML. For SIML-GCN, we construct a Graph Convolution Network (GCN) based on ratings and impose the scale-invariant margin loss on embeddings of users and items. 

For two state-of-the-art metric-learning-based recommendations, CML and SML, SML achieves better performances than CML in most cases due to two drawbacks that existed in CML. But the phenomena is not consistent when evaluated on these three datasets. The performance of SML on Movies is significantly better than on CDs and Books because CDs and Books are more sparse than Movies. SML learns adaptive margins for each user and item, which easily leads to overfitting when the data is sparse.

Besides, Figure~\ref{fig:cmp-real} and Table~\ref{tab:cmp-real} also show that the performances of SIML are evidently superior to the pairwise ranking-based recommendation BPR, CPR. Because BPR and CPR are MF-based methods that violate the triangle inequality according to \cite{shrivastava2014asymmetric}, which leads to the failure of capturing more preference information. Therefore, metric learning with a scale-invariant margin provides accurate top-$k$ personalized recommendations more effectively than pairwise ranking-based models. 

By comparing with LightGCN, LightGCN can not only capture the first-order user-user relation, and user-item relation but also capture the second-order relations. That's why LightGCN performs better than the state-of-the-art metric learning methods and pairwise ranking-based methods. LightGCN can achieve comparable performance to SIML. By formulating the proposed metric learning with GCN, i.e., SIML-GCN, the performance of the proposed method is much better than LightGCN.

\begin{figure}[!tb]
	\centering
	\includegraphics[width=1\linewidth]{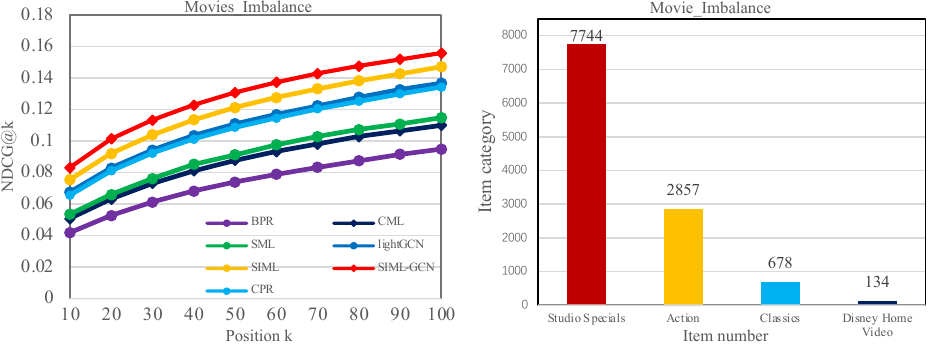}
	\vspace{-18pt}
	\caption{The performance of the proposed SIML on imbalanced data selected on Movies under NDCG@k.}\label{fig:cmp-imb}
%	\vspace{-8pt}
\end{figure}

\paragraph{Efficiency Comparison with Continuous Methods (RQ3) }
As discussed in hashing-based baselines \cite{zhang2016discrete,zhang2017discrete,liu2019compositional}, DSIML maps users and items in $d-$dimension Hamming space, and so the online recommendation is significantly accelerated with the fast similarity search by XOR bit operations \cite{zhang2016discrete}. Table \ref{tab:dsml-eff} shows the efficiency performance of the online recommendation on three datasets by comparing it with the continuous method BPR. Due to continuous methods obtaining real-valued representations, continuous methods perform similarly for the online recommendation stage, and thus we choose the representative BPR for comparison. Table \ref{tab:dsml-eff} shows the significant superiority of the hashing-based method DSIML for providing efficient online recommendations.
\begin{table}%[!t]
	%\extrarowheight=3pt
		\vspace{-4pt}
	\caption{The Effectiveness of Speeding up the Online Recommendation}
	\vspace{-10pt}
	\small
	\centering
	%\begin{tabular*}{0.41\paperwidth}{@{\extracolsep{\fill}}ccccccc}
	\begin{tabular}{c|c|c|c}
		\toprule
		%  \hline
		Datasets & BPR(Seconds)  & DSIML(Seconds) &Speedup ratio \\
		\hline
		Yelp &  2721.84& 213.18&$12.76$ times $\uparrow$\\
		\hline
		Books& 9082.41 & 512.26 &$17.73$ times $\uparrow$ \\
		\hline
		Movies & 41.65 & 5.31&$7.84$ times $\uparrow$\\
		\hline		
		MovieLens &  16.83 & 3.77 &$4.46$ times $\uparrow$\\
		\toprule
	\end{tabular}
	\label{tab:dsml-eff}
%	\vspace{-8pt}
\end{table}
\paragraph{The Effectiveness of Scale-invariant Margin for Imbalanced Data (RQ4) }
Figure \ref{fig:cmp-imb} shows the effectiveness of the proposed scale-invariant margin for items from imbalanced classes. we choose four categories of movies on Movies for evaluation, as shown in the right figure. Experimental results on these imbalanced items validate that the proposed scale-invariant margin is more effective in providing accurate recommendations than the fixed margin used in CML and SML.

\begin{figure}[!t]
	\centering
	\includegraphics[width=1\linewidth]{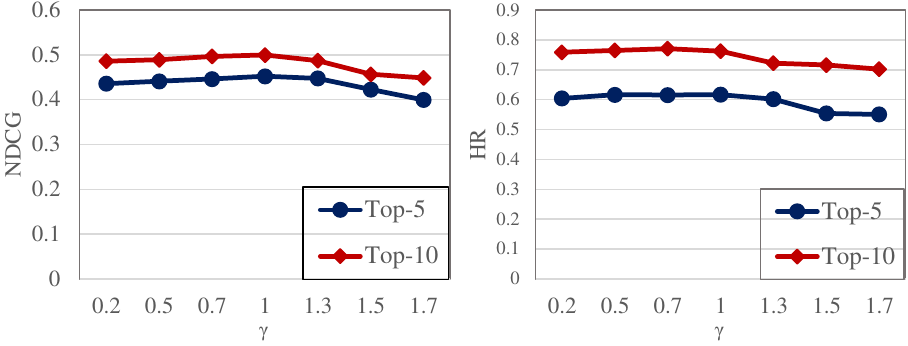}
	\vspace{-16pt}
	\caption{The effect of scale-invariant margin $\gamma$ of SIML for recommendation performances on Movies.}\label{fig:dsml-gamma}
	\vspace{-8pt}
\end{figure}

Figure \ref{fig:dsml-gamma} shows how does the scale-invariant margin $\gamma = \tan{\beta}$ affect the recommendation performance of SIML. we conclude that greater margin $\gamma$ leads to inferior performance under both two metrics, which is reasonable because greater margin leads to greater angle $\alpha$ at the negative sample point in Figure \ref{fig:dsml-margin}, so the recommendation model do not capture fine grained preference difference between negative items and positive items, which leads to poor performance in recommendations. 

A smaller margin leads to better performance because it enforces a stronger margin to the angle $\alpha$. However, the performance is stable when the $\gamma$ increases from 1.3 to 1.7 because a stronger margin makes the optimization problem intractable. Thus the model can only find a locally optimal solution. So we set $\gamma=1$ on the Movies dataset in the experiments. 

%\begin{figure}[!tb]
%	\centering
%	\includegraphics[width=0.6\linewidth]{imb_data}
%	%\vspace{-6pt}
%	\caption{The imbalanced classes selected on Movies.}\label{fig:cmp-data}
%	%\vspace{-10pt}
%\end{figure}

\paragraph{The Influence of Hyper-parameter $\lambda$(RQ5)}

\begin{figure}[!t]
	\centering
	\includegraphics[width=0.95\linewidth]{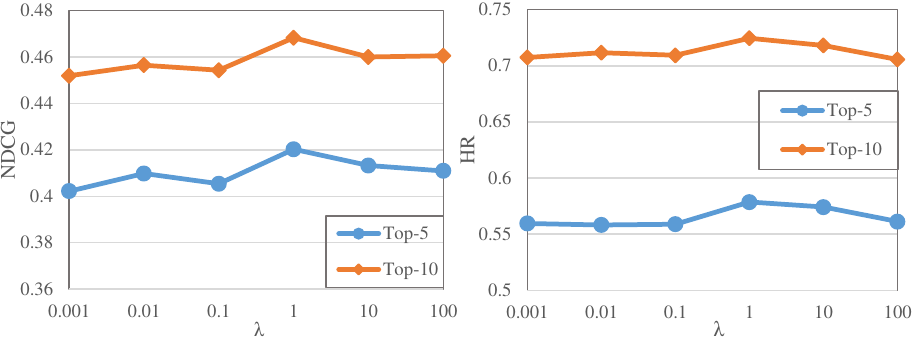}
		\vspace{-10pt}
	\caption{The influence of hyper parameter $\lambda$ of SIML for recommendation performances on Movies.}\label{fig:dsml-lamda}
	\vspace{-10pt}
\end{figure}

Figure \ref{fig:dsml-lamda} displays the result of tuning the hyper-parameter $\lambda$ within the range of $\{e^{-3}, \cdots, e^{2}\}$ by the grid search. we conclude that the performance of recommendations is relatively stable. To achieve better top-$k$ recommendation performance, we choose $\lambda=1$ on the Movies dataset because it achieves the best performance. 

\section{Conclusion}
In this paper, we propose a discrete scale-invariant metric learning (DSIML) for recommendations. Specifically, we first present a scale-invariant margin on angles at negative points and derive a scale-invariant triple hinge loss based on the angle margin. By optimizing the objective, negative items are dragged away from both users and positive items. Besides, the angle margin is invariant with different unbalanced classes. Finally, we formulate the proposed model by incorporating the scale-invariant loss with a pairwise ranking loss and imposing binary constraints on users and items to learn hash codes in Hamming subspace. Experiments on DSIML, its continuous version SIML, and its variation SIML-GCN show that the proposed scale-invariant metric learning respectively outperforms hashing-based baselines, metric learning methods, and the competitive LightGCN.
\balance
\bibliographystyle{ACM-Reference-Format}
%\bibliography{sample-bibliography}
\bibliography{wsdm23}
%\bibliographystyle{ACM-Reference-Format}
%\bibliography{ref} 
\end{document}